# RSAVS superconductors: Materials with a superconducting state that is robust against large volume shrinkage

Cheng Huang,[1,4] Jing Guo,[1,5] Jianfeng Zhang,[2] Karoline Stolze,[3] Shu Cai,[1]
Kai Liu,[2] Hongming Weng,[1,4] Zhongyi Lu,[2] Qi Wu,[1]
Tao Xiang,[1,4] Robert J. Cava,[3] and Liling Sun[1,4,5,*]

[1]Institute of Physics and Beijing National Laboratory for Condensed Matter Physics, Chinese Academy of Sciences, Beijing 100190, China

[2] Department of Physics and Beijing Key Laboratory of Opto-electronic Functional Materials & Micro-nano Devices, Renmin University of China, Beijing 100872, China

[3]Department of Chemistry, Princeton University, Princeton, New Jersey 08544, USA

[4]University of Chinese Academy of Sciences, Beijing 100190, China

[5]Songshan Lake Materials Laboratory, Dongguan, Guangdong 523808, China

The transition temperature ($T_C$) between normal and superconducting states usually exhibits a dramatic increase or decrease with increasing applied pressure. Here we present, in contrast, a type of superconductor that exhibits the exotic feature that $T_C$ is robust against large volume shrinkages (RSAVS, so naming them "RSAVS superconductors") induced by applied pressure. Extraordinarily, our previous studies found that the $T_C$ in the two materials stays almost constant over a large pressure range, e.g. over 136 GPa in the $(TaNb)_{0.67}(HfZrTi)_{0.33}$ high-entropy alloy and 141 GPa in the NbTi commercial alloy. Here，we show that the RSAVS behavior also exists in another high entropy alloy, $(ScZrNbTa)_{0.6}(RhPd)_{0.4}$, and in superconducting elemental Ta and Nb, indicating that this behavior occurs universally in a certain kind of superconductors, composed of only transition metal elements, with a body-centered cubic lattice. Our electronic structure calculations indicate that in the RSAVS state the contribution of the degenerate $d_{x^2-y^2}$ and $d_{z^2}$ orbital electrons remains almost unchanged at the Fermi level, suggesting that these are the electrons that may play a crucial role in stabilizing the $T_C$ in the RSAVS state. We preliminarily analyzed the reasonability and validity of this suggestion by the Homes law.

The proposed new kind of superconductors share universal high-pressure behavior [1-8]; i.e. the superconducting transition temperature ($T_C$) saturates at a critical pressure ($P_C$), the value of which is different for different superconductors, as shown in Fig. 1(a), Fig. S1 in the Supplemental Material [9] (also see [10-26]), and in Table I. These interesting findings have stimulated further investigations [22,27]. Among these superconductors, we find that elemental Ta and Nb also display the same behavior- robust superconductivity against large volume shrinkage (RSAVS), with the primary difference from the alloys being that their $P_C$s start at ambient pressure. Therefore, elemental Nb and Ta can be considered as the simplest, intrinsic RSAVS superconductors, providing a unique opportunity for investigating the nature of this kind of novel superconductivity through ambient-pressure experimental measurements.

Chemically, all these superconductors are composed of only transition metals and, structurally, all have a simple body-centered cubic (bcc) lattice [6-8,28] (Fig. 1(b) and Fig. S2 (see the Supplemental Material [9])). However, what the common feature of the electronic structure is for the exotic RSAVS state and what electrons are responsible for stabilizing the $T_C$ plateau are not clear. In this study, we performed first-principles calculations on the electronic structures for the RSAVS superconductors – Ta and Nb elemental metals, NbTi binary alloy and $(TaNb)_{0.67}(HfZrTi)_{0.33}$ high entropy alloy, by using the projector-augmented wave (PAW) method (see the Supplemental Material [9]) to investigate these issues. As shown in Figs. S3(a)-(d) (see the Supplemental Material [9]), we find that the electronic density of states (DOS) at Fermi level for the Ta and Nb metals and the NbTi alloy is dominated by the $d_{xy}$, $d_{xz}$, and $d_{yz}$ orbitals in the pressure range investigated, with a secondary contribution from the $d_{x^2-y^2}$ and $d_{z^2}$ orbitals. The DOSs of the $p$ and $s$ orbitals are much smaller than those of the $d$ orbitals. Similar behavior is also obtained from the calculations on the $(TaNb)_{0.67}(HfZrTi)_{0.33}$ high-entropy alloy (Fig. S4; see the Supplemental Material [9]). Our results show that the $d_{xy}$, $d_{xz}$, and $d_{yz}$ orbitals and the $d_{x^2-y^2}$ and $d_{z^2}$ orbitals are perfectly degenerate in these RSAVS superconductors, indicating that the results obtained reasonably reflect the electronic signature of the transition metal or alloy with a bcc lattice.

Consequently, we extract the pressure dependence of the partial DOSs for the elemental Ta and Nb (Figs. 2(a) and 2(b); see the Supplemental Material [9]), for the Nb and Ti atoms in the NbTi alloy (Figs. 2(c) and 2(d); see the Supplemental Material [9]), and for the five elements in the $(TaNb)_{0.67}(HfZrTi)_{0.33}$ high-entropy alloy (Figs. 2(e)-2(i); see the Supplemental Material [9]). We find that in all the investigated superconductors the partial DOSs contributed by the $d_{xy}$, $d_{xz}$, and $d_{yz}$ orbitals display a continuous decrease upon compression over the measured pressure range, reflecting the common picture that the DOS decreases under pressure due to the pressure-induced broadening of energy bands in these materials (Fig. S5; see the Supplemental Material [9]). However, it is found that the partial DOSs of the $d_{x^2-y^2}$ and $d_{z^2}$ orbital electrons in the intrinsic RSAVS superconductor Nb or Ta remain almost unchanged with increasing pressure until a distinguishable change above a certain pressure (at 46 GPa for Nb and 50 GPa for Ta, respectively), implying that the superconductors above those pressures leave their *RSAVS* states (Figs. 1(a), 2(a), and 2(b)). For the pressure-induced RSAVS superconductor NbTi alloy and the $(TaNb)_{0.67}(HfZrTi)_{0.33}$ high-entropy alloy, analogous to the behavior of elemental Nb and Ta, the partial DOSs of the $d_{xy}$, $d_{xz}$, and $d_{yz}$ orbital electrons of the Nb or Ti atoms in the NbTi alloy and the Ta, Nb, Hf, Zr, and Ti atoms in the high-entropy alloy also decrease in the pressure regimes of the RSAVS state, but the partial DOSs contributed by the $d_{x^2-y^2}$ and $d_{z^2}$ orbital electrons remain almost unchanged (Figs. 2(c) and 2(d), Figs. 2(e)-2(i)). These consistent results obtained in these different superconductors from the simple elements to the complicated high-entropy alloy suggest that the $d_{x^2-y^2}$ and $d_{z^2}$ orbital electrons are closely connected to the RSAVS state, which provides a unique case that some peculiar electronic orbitals are vital in determining the $T_C$ value [29]. For the partial DOSs contributed by the $p$ and $s$ orbitals, no feature is concurrent with the $T_C$ plateau found in the RSAVS state.

According to the BCS theory [30], $T_C$ is controlled by density of state at the Fermi level and attractive interaction associated with phonon energy. However, what the relations of the partial DOSs contributed by the $d_{x^2-y^2}$ and $d_{z^2}$ orbital electrons with the above variables are and whether the stable partial DOSs against pressure is truly essential for stabilizing the RSAVS state deserve further experimental and theoretical investigations. Therefore, we turn to use the Homes law [31,32] to have a preliminary analysis on the reasonability and validity of our results. From the law ($\rho_s = A\sigma T_C$), it is known that the superconducting transition temperature ($T_C$) is determined by two quantities, superfluid density ($\rho_s$) and conductivity ($\sigma$), which is the inverse of resistivity, in the normal state at $T_C$, as well as a universal constant ($A$), which has a value about 120 ±25. Assuming that the superconducting carrier density (superfluid density $\rho_s$) in the RSAVS superconductors is contributed by the electrons from the $d_{x^2-y^2}$ and $d_{z^2}$ orbitals, we thus employ the Homes law to test our assumption for the elemental Ta and Nb (see the Supplemental Material [9]). We find that our data are all in the line with slope of unity (see the balls in Fig. 3); in particular, the data of elemental Ta obtained at 7.3 GPa also excellently obeys Homes law (see the blue ball in Fig. 3), implying that the electrons contributed by the $d_{x^2-y^2}$ and $d_{z^2}$ orbitals are the superconducting carriers. We attempted to fit the data of the compressed Ta at higher pressure but failed due to the pressure-induced high ductility of the sample that prevents us to obtain reliable resistivity. Moreover, we also take the electrons contributed by the $d_{xy}$, $d_{xz}$, and $d_{yz}$ orbitals, which are the electrons dominating the Fermi surface of the RSAVS states of Ta and Nb, as the $\rho_s$ in the fitting of the Homes law, and find that they obviously deviate from the line (see the stars in Fig. 3). These results exclude the possibility that the electrons from the $d_{xy}$, $d_{xz}$, and $d_{yz}$ orbitals are the superconducting carriers of the RSAVS superconductors.

The behavior of this anomalous RSAVS state found in the superconductors from the simple transition metals to the complicated alloys is distinct from that of other conventional or unconventional superconductors: the $T_C$s of copper-oxide or iron-pnictide superconductors are highly sensitive to the external pressure [1-5,33-36], for example, as are the $T_C$s of conventional superconductors in general [37-39]. Therefore, this exotic kind of superconducting state provides not only a unique platform for investigating the correlations of superconductivity with chemical constitution and crystal and electronic structures for conventional superconductors but also fresh information for achieving a better understanding of the superconductivity in the conventional or unconventional superconductors based on transition metals.

**Acknowledgements**

We thank Prof. Jiangping Hu and Jiacheng Gao for useful discussions. The work in China was supported by the National Key Research and Development Program of

China (Grant No. 2017YFA0302900, 2016YFA0300300 and 2017YFA0303103), the NSF of China (Grant Numbers 11604376, 11774422, 11774424), and the Strategic Priority Research Program (B) of the Chinese Academy of Sciences (Grant No. XDB25000000). J. G. is grateful for support from the Youth Innovation Promotion Association of the CAS (2019008). We also acknowledge support from the K. C. Wong Education Foundation (GJTD-2018-01), Beijing Municipal Science & Technology Commission (Z181100004218001) and Beijing Natural Science Foundation (Z180008). The work at Princeton was supported by the Gordon and Betty Moore Foundation EPiQS initiative, Grant GBMF-4412.

*Correspondence and requests for materials should be addressed to L.S (llsun@iphy.ac.cn).

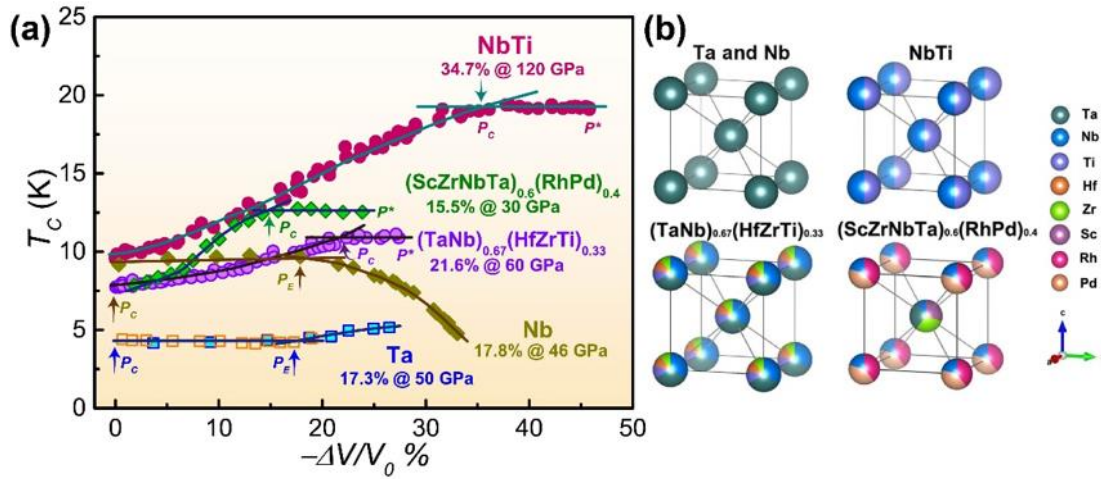

**FIG. 1. Superconductivity and crystal structure for the *RSAVS* superconductors**

(a) The pressure-dependent change in the superconducting transition temperature ($T_C$) of the (TaNb)$_{0.67}$(HfZrTi)$_{0.33}$ and (ScZrNbTa)$_{0.6}$(RhPd)$_{0.4}$ high entropy alloys, the NbTi alloy, and the elemental Ta and Nb. In order to facilitate the comparison of the different materials, we use the volume shrinkage ($-\Delta V/V$) as a variable. Arrows in the diagram indicate the critical pressure ($P_C$) where the RSAVS state emerges. $P_C$ is about 30 GPa [the corresponding volume ($-\Delta V/V_0$) change is about 15.5%] for the (ScZrNbTa)$_{0.6}$(RhPd)$_{0.4}$ superconductor, 60 GPa ($-\Delta V/V_0$=21.6%) for the (TaNb)$_{0.67}$(HfZrTi)$_{0.33}$ superconductor and 120 GPa ($-\Delta V/V_0$=34.7%) for the NbTi superconductor, while $P_C$ is 1 bar for the elemental Ta and Nb superconductors. $P_E$ and $P^*$ represent the end pressure of the RSAVS state and the highest pressure measured for the RSAVS state, respectively. (b) Sketches for the lattice structure of the (TaNb)$_{0.67}$(HfZrTi)$_{0.33}$ and (ScZrNbTa)$_{0.6}$(RhPd)$_{0.4}$ high entropy alloys, NbTi alloy, elemental Ta and Nb, which all possess the body-centered cubic structure.

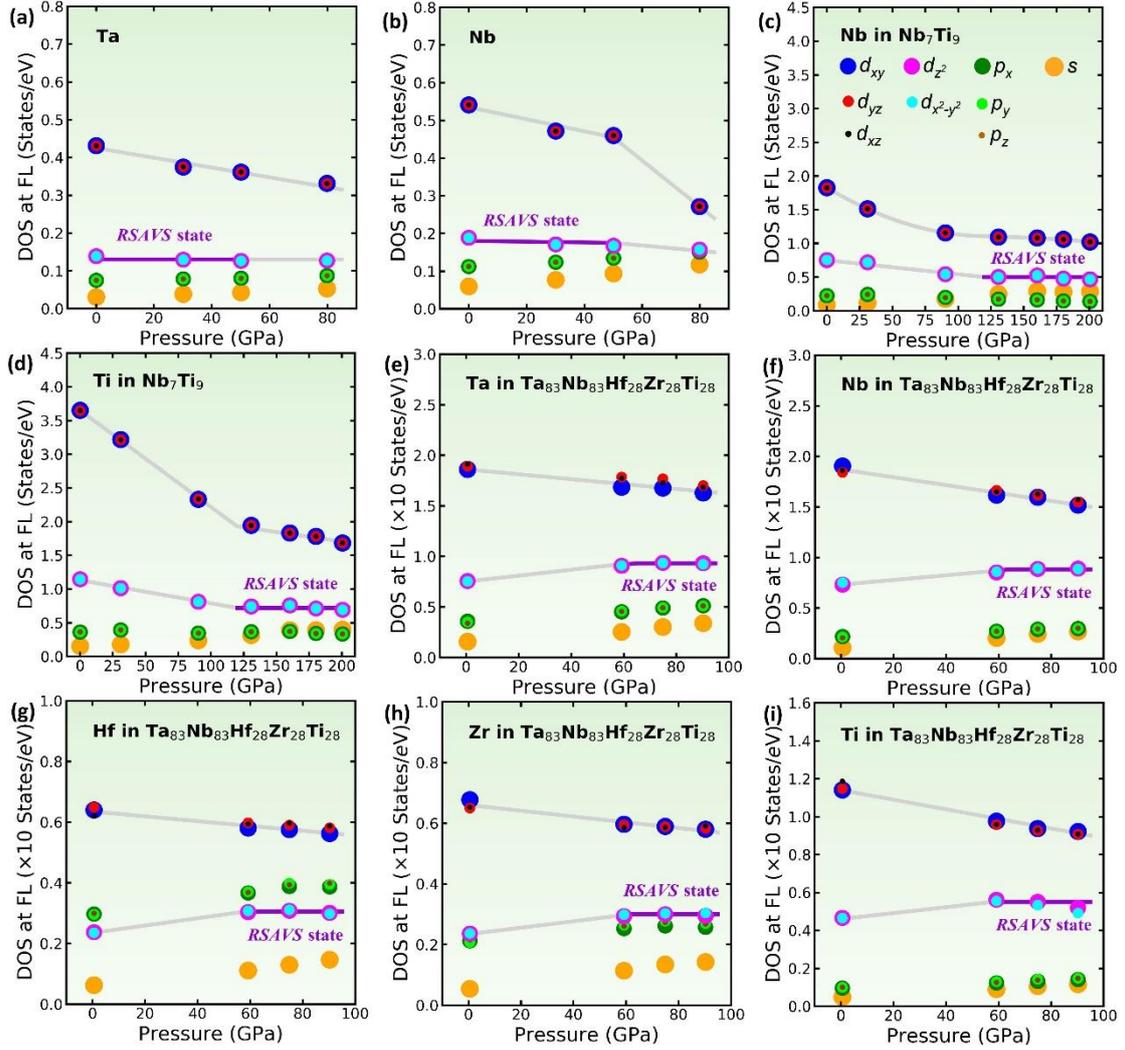

**FIG. 2. The calculated pressure dependence of the density of states (DOSs) at the Fermi level (FL) contributed by different orbitals in the RSAVS superconductors.** (a-b) Partial DOSs at the FL as a function of pressure for the elemental Ta and Nb. (c-d) Pressure dependence of the partial DOSs at the FL for the Nb and Ti, respectively, in the NbTi alloy. (e-i) Plots of partial DOSs at the FL versus pressure for the Ta, Nb, Hf, Zr and Ti in the $(TaNb)_{0.67}(HfZrTi)_{0.33}$ high-entropy alloy. Although a 5×5×5 supercell, which involves 250 atoms and reaches the limit state of our computing, is adopted for the calculations on the high entropy alloy, it seems still not enough for achieving a

perfect degeneracy for the $d$ orbitals. The trend of the DOSs versus pressure, however, is analogous to those obtained in elemental Ta and Nb as well as NbTi alloy – the DOSs contributed by $d_{x^2-y^2}$ and $d_{z^2}$ orbital electrons remain almost unchanged in the pressure regimes of the RSAVS state.

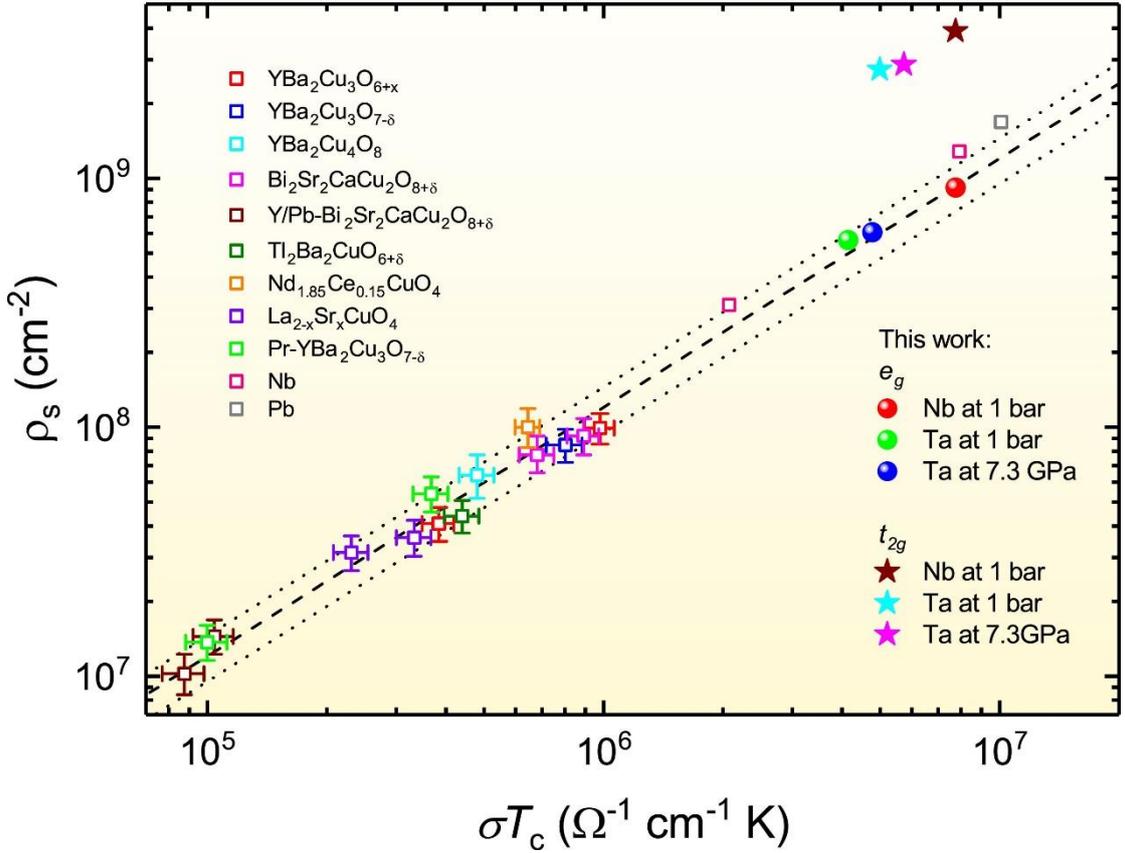

**FIG. 3. The superfluid density ($\rho_s$) as a function of the product of the conductivity ($\sigma$) and the superconducting transition temperature ($T_C$) for a variety of copper oxides and elemental metals.** The data denoted by open squares are taken from the Ref. [31]. The data presented by solid balls are our results, in which the $\rho_s$ values are derived from the partial DOSs contributed by the $d_{x^2-y^2}$ and $d_{z^2}$ orbitals. The data presented by stars are also our results, in which the $\rho_s$ values are derived from the partial

DOSs contributed by the $d_{xy}$, $d_{xz}$, and $d_{yz}$ orbitals. The $e_g$ represents $d_{x^2-y^2}$ and $d_{z^2}$ orbitals, the $t_{2g}$ stands for $d_{xy}$, $d_{xz}$, and $d_{yz}$ orbitals.

**Table I** Overview with superconducting transition temperature $T_C$, critical pressure and volume change of the high-entropy alloy, NbTi alloy and elemental Ta and Nb superconductors. $T_{C\text{-am}}$ and $T_{C\text{-pl}}$ represent the $T_C$ at ambient pressure and the $T_C$ that shows plateau behavior (the RSAVS state). $P_C$, $P^*$, and $P_E$ stand for the critical pressure where the RSAVS state emerges, the experimentally highest pressure measured for the RSAVS state, and the end pressure of the RSAVS state. $\Delta P$ and $-\Delta V/V_C$ represent the range of pressure and corresponding volume change in the RSAVS state.

| Materials | $T_{C\text{-am}}$ | $T_{C\text{-pl}}$ | $P_C$ | $P^*$ | $P_E$ | $\Delta P$ | $-\Delta V/V_C$ (%) |
|---|---|---|---|---|---|---|---|
| (TaNb)$_{0.67}$(HfZrTi)$_{0.33}$ (Ref. [6]) | 7.7 K | ~10 K | 60 GPa | 190 GPa | - | ≥ 130 GPa | ≥ 6.8 |
| (ScZrNbTa)$_{0.6}$(RhPd)$_{0.4}$ (this study) | 7.3 K | ~12.5 K | 30 GPa | 72 GPa | - | ≥ 42 GPa | ≥ 9.2 |
| NbTi (Ref. [7]) | 9.6 K | ~19 K | 120 GPa | 262 GPa | - | ≥ 142 GPa | ≥ 14.4 |
| Ta (Ref. [8]) | 4.2 K | ~4.2 K | 1 bar | 96 GPa | 50 GPa | 50 GPa | 17.3 |
| Nb (Ref. [8]) | 9.1 K | ~9.1 K | 1 bar | 132 GPa | 46 GPa | 46 GPa | 17.8 |